\documentclass[iop]{emulateapj}

\usepackage{float}
\usepackage[normalem]{ulem}

\begin{document}

\title{The Generation of the Distant Kuiper Belt by Planet Nine\\ from an Initially Broad Perihelion Distribution}

\author{Tali Khain\altaffilmark{1}}
\author{Konstantin Batygin\altaffilmark{2}}
\author{Michael E. Brown\altaffilmark{2}}

\altaffiltext{1}{Department of Physics and Department of Mathematics, University of Michigan, Ann Arbor, MI 48109, USA.}
\altaffiltext{2}{Division of Geological and Planetary Sciences, California Institute of Technology, Pasadena, CA 91125, USA}

\email{talikh@umich.edu}

\begin{abstract}
The observation that the orbits of long-period Kuiper Belt objects are anomalously clustered in physical space has recently prompted the Planet Nine hypothesis - the proposed existence of a distant and eccentric planetary member of our solar system. Within the framework of this model, a Neptune-like perturber sculpts the orbital distribution of distant Kuiper Belt objects through a complex interplay of resonant and secular effects, such that in addition to perihelion-circulating objects, the surviving orbits get organized into apsidally aligned and anti-aligned configurations with respect to Planet Nine's orbit. In this work, we investigate the role of Kuiper Belt initial conditions on the evolution of the outer solar system using numerical simulations. Intriguingly, we find that the final perihelion distance distribution depends strongly on the primordial state of the system, and demonstrate that a bimodal structure corresponding to the existence of both aligned and anti-aligned clusters is only reproduced if the initial perihelion distribution is assumed to extend well beyond $\sim 36$ AU. The bimodality in the final perihelion distance distribution is due to the existence of permanently stable objects, with the lower perihelion peak corresponding to the anti-aligned orbits and the higher perihelion peak corresponding to the aligned orbits. We identify the mechanisms which enable the persistent stability of these objects and locate the regions of phase space in which they reside. The obtained results contextualize the Planet Nine hypothesis within the broader narrative of solar system formation, and offer further insight into the observational search for Planet Nine.
\end{abstract}

\section{Introduction}

The continued unveiling of the outer solar system by astronomical surveys has revealed a puzzling orbital clustering among the distant Kuiper Belt objects (KBOs). \citet{trujillo} were the first to note a peculiar grouping in the argument of perihelion of long-period KBOs. Subsequently, \citet{batygin_evidence} found that the orbits of KBOs with semi-major axes $a \gtrsim 250$ AU and perihelion distances $q > 30$ AU are in fact anomalously aligned in physical space i.e., these objects exhibit statistically significant clustering in longitude of perihelion as well as ascending node. To explain this persistent orbital grouping, \citet{batygin_evidence} proposed the existence of a Neptune-like perturber - ``Planet Nine" - that is apsidally anti-aligned with the aforementioned KBOs, and generates their orbital confinement through a complex interplay of resonant and secular effects \citep{Becker, BatMorb}.

While this orbital clustering constitutes the most visually striking line of evidence for the existence of Planet Nine \citep{brown_constraints}, it is by no means the only one. In particular, the detachment of high perihelion ($q \gtrsim 36$ AU) KBOs from Neptune requires an extra gravitational influence, which a distant perturber can successfully provide. In addition, Planet Nine naturally explains the dynamics of highly inclined large semi-major axis centaurs \citep{high-a_centaurs,batygin_evidence}, as well as retrograde KBOs with $a < 100$ AU, such as Drac and Niku \citep{high_inc}. Lastly, the presence of an inclined distant planet in our solar system and the associated secular torque can explain the otherwise mysterious six-degree obliquity of the Sun \citep{Bailey, Lai, Gomes}.

Given the inferred distant and eccentric orbit of Planet Nine, a natural question that arises concerns its formation. To this end, three main mechanisms have been considered in the literature. In the first scenario, Planet Nine is captured into its current orbit, either from another solar system, or as a free-floating planet \citep{capture, P9_cross_sections}. In the second, the formation of Planet Nine is a two-step process. Particularly, originating in the inner solar system, Planet Nine is envisioned to have been scattered out to a wide orbit by the giant planets (possibly during the transient period of dynamical instability prescribed by the Nice model: \citealt{Nice_1, Nesvorny2011, BatBrownBetts}). Since the hypothesized current orbit of Planet Nine has a high perihelion ($q_9 > 250$ AU), gravitational influence of the solar system's birth cluster \citep{MorbyLevison, AdamsReview} or dynamical friction from an extended massive planetesimal disk \citep{disk} is then invoked to perturb the orbit and raise Planet Nine's perihelion distance. The third mechanism is in-situ formation, that is, the formation of Planet Nine in its current position \citep{Bromley}. 

Given how distant the orbit of Planet Nine is hypothesized to be, the third scenario would require efficient conglomeration of solids at a large stellocentric distance as well as a disk that extends to hundreds of AU. The probability of the other two mechanisms was evaluated by \citet{P9_cross_sections} and \citet{disk}; the authors show that in both cases, special conditions are required to reproduce and retain the orbit of Planet Nine. In light of these results, the formation mechanism of Planet Nine remains unclear and of considerable interest.

Unlike the in-situ formation scenario,
we note that both the capture and outward scattering mechanisms rely on the occurrence of stellar encounters. Importantly, the circumstances of these stellar interactions certainly affect the initial Kuiper Belt structure. In other words, the same perturbations that would have lifted Planet Nine's perihelion would have also emplaced a population of distant KBOs onto high-$q$ orbits. As of yet, however, a systematic study of the evolution of the Kuiper Belt from different initial distributions has not been done. 

In this paper, we investigate this question in detail. The paper is structured as follows. In section \ref{sec:sims}, we introduce our simplified numerical model and outline the key differences in the evolution of the narrow and broad-$q$ Kuiper Belt populations. In particular, we focus on the dynamics of the objects in the apsidally ``aligned" and ``anti-aligned" classes (with respect to Planet Nine's longitude of perihelion). We then contextualize our results within the framework of more realistic N-body simulations. We summarize and discuss the implications of our results in section \ref{sec:discussion}, in particular focusing on the extent to which our model can aid in interpreting current observations.

\vspace{5mm}

\section{Numerical Simulations}
\label{sec:sims}

To study the dependence of the Planet Nine hypothesis on the initial conditions of the Kuiper Belt, we explore the evolution of two initially distinct Kuiper Belt distributions. Our choice of distributions is motivated by the proposed mechanisms for Planet Nine's formation. The first population, the ``narrow'' Kuiper Belt ($q \in [30, 36]$), corresponds to the conventional definition of the scattering disc \citep{Gladman}. The second ``broad'' population ($q \in [30, 300]$), is associated with mechanisms involving stellar encounters, which could lift the perihelion distance of both Planet Nine and the KBOs. In the scattering or capture scenarios, then, the Kuiper Belt $q$-distribution may range from Neptune's semi-major axis all the way to Planet Nine's perihelion distance and further.

With these considerations in mind, we begin our study of the evolution of these two populations by employing an idealized numerical model. We carry out two sets of N-body simulations that span $4$ Gyr of integration time, and utilize the \texttt{mercury6} software package \citep{mercury}, employing the hybrid Wisdom-Holman/Bulirsch-Stoer algorithm. 

As a first step, we consider a simplified planar model of the solar system, which includes Neptune and Planet Nine as active bodies, but replaces Jupiter, Saturn, and Uranus with a solar $J_2$ moment as in \citet{batygin_evidence}. That is, we account for the gravitational potential of the three planets by adding a non-zero quadrupolar field to the Sun, and extending its radius to Uranus' semi-major axis. In this semi-averaged model, we introduce a 10 $m_{\oplus}$ Planet Nine with semi-major axis $a_9 = 700$ AU, eccentricity $e_9 = 0.6$, and inclination $i_9 = 0^{\circ}$, and set the timestep to $16$ years (one-tenth of Neptune's orbital period). 

While capture of Planet Nine by the solar system is unlikely to impulsively perturb the Kuiper belt in any appreciable manner, the same is not necessarily true if Planet Nine attained its initial long-period orbit by scattering off of Jupiter or Saturn. Simulations of this process, however \citep{fiveplanetBBB}, demonstrate that even in this violent scenario, the dynamical footprint of the scattering upon the Kuiper belt is modest in comparison with that entailed by the Nice model instability \citep{levison}.

The first set of numerical simulations contains a synthetic distant Kuiper Belt in which the perihelion distance of the test particles is uniform in the range $q \in [30, 36]$ AU (the ``narrow" Kuiper Belt), consistent with the expectations of the Nice model in the absence of external perturbations. Conversely, in our second set of simulations, the perihelion distance of the test particles uniformly ranges between $q \in [30, 300]$ AU (the ``broad" Kuiper Belt). The initial distributions of the other orbital parameters of the objects are the same in both experiments: the semi-major axes are drawn uniformly\footnote{In the broad distribution, all generated objects with $q > a$ are discarded from the simulations; in the allowed regions, the $a$ and $q$ distributions are uniform.} from $a \in [150, 1000]$ AU, inclinations follow a half-normal distribution centered at $i = 0^\circ$ with $\sigma_i = 5^\circ$, and the argument of perihelion ($\omega$), longitude of ascending node ($\Omega$), and mean anomaly ($M$) are uniformly drawn from the full $[0, 360)$ degree range. Each set of simulations contains about $10,000$ massless particles.

\begin{figure}[t!]
\epsscale{1}
  \begin{center}
      \leavevmode
\includegraphics[width=85mm]{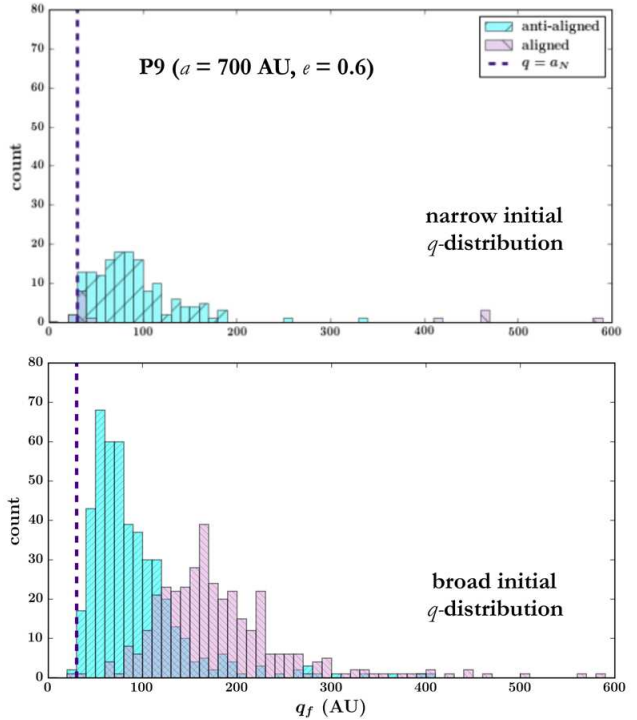}
\caption{Final $q$-distribution of librating aligned and anti-aligned objects surviving the full $4$ Gyr integration. The top panel shows objects with initial $q_0 \in [30, 36]$ AU, while the bottom panel shows objects with initial $q_0 \in [30, 300]$ AU. Although the ``narrow" Kuiper Belt is severely depleted (top panel), the distribution of the anti-aligned objects is similar in the top and bottom panel. In contrast, only the ``broad" Kuiper Belt retains the aligned objects, which must be in the high-$q$ regime in order to survive.} 
\label{fig:qdist}
\end{center}
\end{figure}

\begin{figure}[t!]
\epsscale{1}
  \begin{center}
      \leavevmode
\includegraphics[width=85mm]{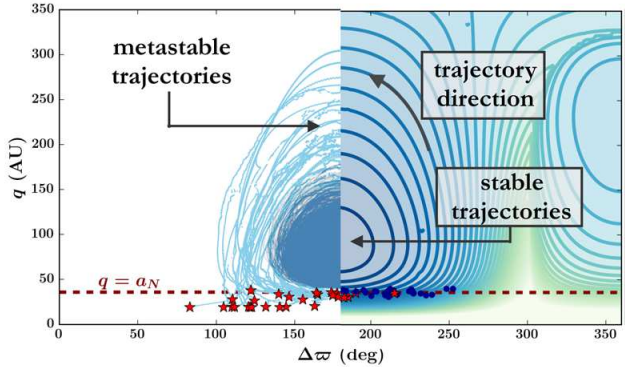}
\caption{The $q-\Delta \varpi$ phase plane, showing the trajectories of a subset of the anti-aligned objects (left) and contours of the secular Hamiltonian (right). The test particles follow these secular contours, which were computed as in \citet{BatMorb}. The light blue objects, which attain high-$q$ values, are metastable. Initialized in the region denoted by the blue dots, such objects are first driven to high-$q$ values and then subsequently plunged into the inner solar system due to secular interactions with Planet Nine. The dark blue trajectories, however, experience only mild $q$ and $\Delta \varpi$ librations, and remain stable throughout the span of the integration.}
\label{fig:antialigned}
\end{center}
\end{figure}

Objects surviving the full $4$-Gyr integrations make up just a small fraction of the initially launched population, as can been seen in Figure \ref{fig:qdist}. Rather than analyzing the $q$-distribution of the entire aggregate of test-particles, here we focus on the behavior of specific dynamical classes of KBOs. Particularly, the two dynamical populations of interest consist of the objects which experience longitude of perihelion librations with respect to Planet Nine's longitude of perihelion, notated as $\Delta \varpi$: the ``anti-aligned" and the ``aligned" populations. The objects anti-aligned to Planet Nine are identified as those which librate around $\Delta \varpi = 180^\circ$, while those which are aligned librate around $\Delta \varpi = 0^\circ$. 

In our analysis, we then track the evolution of the initially uniform perihelion distance distribution of the anti-aligned and aligned populations. Notably, we find that the final $q$-distribution of these objects in the narrow and broad Kuiper Belt populations is strikingly different. The top and bottom panels of Figure \ref{fig:qdist} show the final distribution of the narrow and broad populations, respectively. Evidently, very few aligned objects initialized within the narrow population survive for $4$ Gyr (the surviving aligned population makes up only $0.14\%$ of the original $\sim$10,000 objects). On the other hand, a clear bimodal structure - with the higher $q$-peak corresponding to the aligned population ($3.70\%$ surviving), and the lower $q$-peak corresponding to the anti-aligned population ($5.84\%$ surviving) - emerges in the numerical experiments characterized by the initially broad $q$-distribution of KBOs. Of course, not all of these objects are observable; applying a $d^{-4}$ detection bias, we compute that the median final perihelion distance of the detectable objects in the narrow distribution (top panel of Figure \ref{fig:qdist}) is 38.13 AU for the anti-aligned KBOs and 33.07 AU for the aligned KBOs, and in the broad distribution (bottom panel), 44.16 AU for the anti-aligned KBOs, and 100.29 AU for the aligned KBOs.

In the following sections, we now consider the difference between the final perihelion distance distributions of the narrow and broad populations in greater detail.

\begin{figure}[t!]
\epsscale{1}
  \begin{center}
      \leavevmode
\includegraphics[width=85mm, height=50mm]{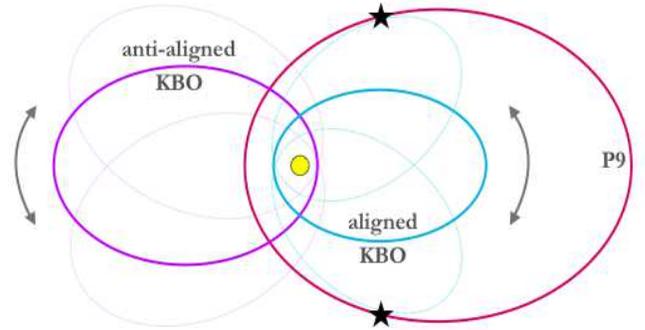}
\caption{A diagram of the geometry of the aligned (blue) and anti-aligned (purple) Kuiper Belt objects in the presence of Planet Nine (pink). The orbit of the aligned object may librate in the offset of longitude of perihelion, but when this libration exceeds a critical angle, the blue and pink orbits cross, and the aligned object is pushed into the collisional domain with Planet Nine. The anti-aligned objects permanently reside in this collision region, and the surviving ones gain their stability through resonance interactions with Planet Nine.} 
\label{fig:geometry}
\end{center}
\end{figure}

\subsection{The Anti-Aligned Population}

As shown by \citet{BatMorb}, the observed dynamics of the anti-aligned population stems from a complex interplay between mean-motion resonances and secular interactions. That is, in face of orbit-crossing, the anti-aligned objects glean long-term stability by chaotically hopping \citep{Becker} between phase-protected mean-motion commensurabilities \citep{malhotra, Millholland}. At the same time, the specific (corotation) nature of these resonances implies that the anti-alignment of the orbits itself is maintained via a phase-averaged flavor of dynamical coupling. Although this behavior is well understood, the effect of the initial $q$-distribution on this population is not immediately evident.

Interestingly, the distribution of the anti-aligned objects (light blue, Figure \ref{fig:qdist}) is similar in both the narrow and broad populations. Yet, we see that these two distributions are not identical: the broad population is able to retain a greater number of anti-aligned objects, as many of these particles start out and remain detached from Neptune and thus experience minimal scattering interactions with the giant planets. Even so, the final $q$-distributions of both populations are dominated by objects with middling perihelion distances, with most of the objects possessing $q_f < 100$ AU. 

\begin{figure*}[t!]
\epsscale{1}
  \begin{center}
      \leavevmode
\includegraphics[width=175mm]{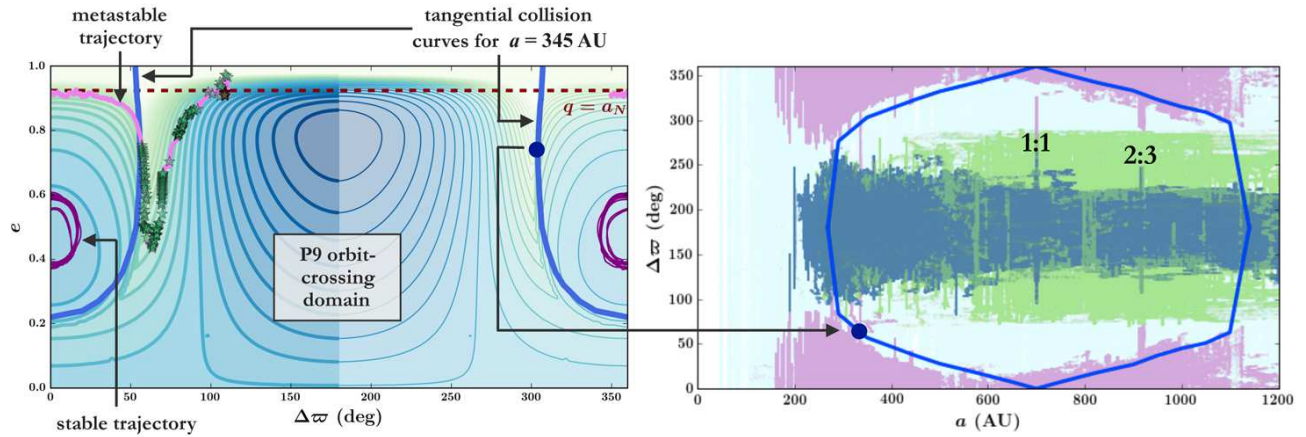}
\caption{The left panel shows the $e-\Delta \varpi$ phase plane with contours of the averaged Hamiltonian in the background for a constant KBO semi-major axis of $a = 345$ AU. The blue boundary shows the tangential collision curve with Planet Nine. The orbits of objects within this boundary do not cross Planet Nine's orbit, and are thus protected from close encounters (see the stable purple libration cycles). 
Objects that do cross this curve, however, are eventually driven into the inner solar system through Planet Nine-induced secular dynamics, becoming short-lived centaurs. An instance of such dynamics is shown in pink, with the green stars denoting close encounters. This object only survives 100 Myr, as it is suffers a rapid decrease in $q$ and is engulfed by the canonical planets shortly after crossing into the Planet Nine collisional domain. The right panel shows the full trajectories of all surviving objects in the planar simulations on the $\Delta \varpi-a$ phase plane. The surviving population includes objects that are aligned (pink), anti-aligned (dark blue), highly-inclined and often retrograde (green), and circulating in $\Delta \varpi$ (light blue). The dynamics of the highly inclined population are outside the scope of this work, but are thoroughly discussed in \citet{BatMorb}. The blue collision curve is constructed by locating the maximal allowed libration width of the relative longitude of perihelion in the corresponding $e-\Delta \varpi$ plot for each value of $a$. All orbital configurations within the blue curve are orbit-crossing with Planet Nine. Thus, all non-resonant aligned objects are found outside of this orbit-crossing region.}
\label{fig:aligned}
\end{center}
\end{figure*}

It is intriguing to note that anti-aligned objects in the broad Kuiper Belt population with initial $q$ in excess of $\sim 100$ AU do not survive the full integration time. This is because such objects reside on Planet Nine-induced secular trajectories that experience high-amplitude perihelion distance oscillations. Once their maximum $q$ is attained, the objects continue to approximately follow the level curves of the averaged Hamiltonian (computed as in \citealt{BatMorb}), and are driven into Neptune-crossing orbits. At this point, these objects typically become centaurs, and as centaurs have short dynamical lifetimes \citep{centaurs_1, centaurs_2}, we allow them to be absorbed by the central body rather than resolving their interactions with the canonical giant planets.

We illustrate this effect in Figure \ref{fig:antialigned}: the light blue trajectories shown in the $q-\Delta \varpi$ phase plane attain high-$q$ values as they librate in $\Delta \varpi$. The initial ($\Delta \varpi_0, q_0$) values of these objects are denoted with blue dots, and the red stars indicate the moment when these objects cross Uranus' orbit. These metastable objects survive less than a single libration in $\Delta \varpi$, as they are driven to low $q$ before returning to their initial locations in the phase plane (see the lower region of Figure \ref{fig:antialigned}). In fact, any real objects currently residing at any point on these light blue secular trajectories are expected to become unstable due to this effect.

In contrast with this metastable evolution, the dark blue objects nested within the lighter blue loops are the trajectories of stable anti-aligned objects. A schematic of the geometry of the longitude of perihelion libration is shown in Figure \ref{fig:geometry}. These particles experience only mild libration in both $q$ and $\Delta \varpi$, and are thus able to remain in a long-term stable configuration protected from Neptune-scattering. Accordingly, we find that after a $4$ Gyr integration, all anti-aligned objects which attain high-$q$ values in their trajectories are eventually ejected. This qualitatively explains why the perihelion distribution of stable anti-aligned objects is limited to $q \lesssim 100$ AU in both simulations of the broad and the narrow Kuiper belt populations.

\vspace{2mm}
\subsection{The Aligned Population}
In addition to the anti-aligned class of KBOs that executes resonant-secular dynamics, the Planet Nine hypothesis also entails a population of apsidally aligned objects. The biggest difference between the synthetic Kuiper Belts created in the two numerical experiments shown in Figure \ref{fig:qdist} lies in the behavior of these objects, whose dynamics are dominated by purely secular effects. Almost no aligned objects survive the full integration time in the narrow Kuiper Belt population; in fact, the few that do survive are found in high order resonances with Neptune. In contrast, the broad Kuiper Belt retains a large population of high-$q$ aligned objects.

Interestingly, the mechanism that depletes the lower-$q$ aligned population in both distributions is a combination of secular dynamics and scattering interactions with Planet Nine and Neptune. To envision this process in physical space, consider the geometry of a typical orbital configuration occupied by a stable particle belonging to the dominantly secular, apsidally aligned population of long-period KBOs (Figure \ref{fig:geometry}). The pink (outer) ellipse represents Planet Nine's orbit, while the blue (inner) ellipse is an example orbit of an apsidally aligned test particle (note that we are operating under the assumption that Planet Nine and the test particle lie in the same plane). Clearly, as the relative longitude of perihelion of the object librates, the inner blue orbit explores some portion of the region enveloped within Planet Nine's pink orbit. Once the orbit of the object crosses the orbit of Planet Nine, however, the object is in danger of experiencing nearly tangential close encounters with Planet Nine.

For a fixed $a < a_9$ of the test particle, there exists a protected region in the $e-\Delta \varpi$ phase plane, in which the object can experience safe librations within the orbit of Planet Nine. If $a > a_9$, there exists a similar protected region in which the orbits of the aligned objects engulf the orbit of Planet Nine without crossing it. In configurations other than these apsidally aligned geometries, however, close encounters must occur.

Due to secular interactions with Planet Nine, these aligned objects execute libration cycles in the $e-\Delta \varpi$ phase plane. Depending on the initial $q$, some of these loops are fully contained within the protected region of phase space, that is, the colinear orbits of these objects remain entirely within or outside of Planet Nine's orbit, and in this case the objects survive the full integration time. An example of such a trajectory is shown in purple in the left panel of Figure \ref{fig:aligned}.

For objects initialized on Neptune-hugging orbits, however, secular evolution drives the particle towards an orbit-crossing configuration with Planet Nine. Near this boundary, secular theory formally breaks down, as close encounters ensue. Intriguingly, our simulations show that despite repeated close encounters, nearly tangential objects are not removed from the system by Planet Nine. Instead, these interactions chaotically shift their positions in the phase plane. As a result, the object crosses the tangential collision curve and wanders into the orbit-crossing regime with Planet Nine. Once the object moves away from the tangential collision curve, secular theory regains its accuracy \citep{Gronchi}: the object finds itself on a relatively smooth secular trajectory that drives it into the inner solar system via a rapid decrease in $q$. An example of such dynamics is shown in pink in the left panel of Figure \ref{fig:aligned}. This transition from an initially stable secular cycle to an unstable one is the primary mechanism which depletes the lower-$q$ population of aligned objects, and qualitatively explains the lack of long-term stable aligned objects in the top panel of Figure \ref{fig:qdist}.

The tangential collision curve (blue boundary in the left panel of Figure \ref{fig:aligned}) and the entire $e-\Delta \varpi$ phase plane presented in this plot are mapped for a fixed value of $a$ (in this case, the semi-major axis of the objects shown is $a = 345$ AU). To compute the collision curve for a particular $a$, we find the value of the eccentricity of the test particle for which its orbit first touches Planet Nine's orbit, given a fixed $\Delta \varpi$ (marked by stars in Figure \ref{fig:geometry}). We then iterate this procedure for varying values of $\Delta \varpi$. The location of the tangential collision curve is thus determined by the semi-major axis of the object. In fact, the area of the non-orbit-crossing region and the width of the allowed longitude of perihelion librations is dependent on the $a/a_9$ ratio. As the semi-major axis of the test particles approaches that of Planet Nine, the region in which the orbits of these objects can remain without crossing Planet Nine's orbit becomes more and more limited, until it vanishes in the limit $a \to a_9$.

Indeed, this effect is evident in our simulation results. In the right panel of Figure \ref{fig:aligned}, we show the $\Delta \varpi-a$ phase plane for all objects that survive the full $4$ Gyr integrations in the planar simulations. The dark blue objects, which librate about $\Delta \varpi = 180^{\circ}$, are the anti-aligned objects. The aligned objects, which librate about $\Delta \varpi = 0^{\circ}$, are denoted in pink. The blue curve overlaid on the simulation results represents the tangential collision region with Planet Nine. To compute this boundary, we construct the appropriate tangential collision curve for a fixed $a$ in the $e-\Delta \varpi$ phase plane, as shown in Figure \ref{fig:aligned}, locate the maximum libration width (denoted with a dark blue dot), and map it onto the $\Delta \varpi-a$ phase plane. 

The regions outside of this collision curve allow for orbital configurations in which orbit crossing does not occur. Within the region bounded by the curve, however, all orbital configurations are marked by orbit crossing with Planet Nine. As we discussed earlier, it is important to note that close encounters with Planet Nine do not directly destabilize these objects. Instead, these interactions simply act to stochastically modulate the KBOs within the Planet Nine orbit-crossing regime. Once in this region, however, objects are carried into the inner solar system via secular trajectories, and are ejected or absorbed by the Sun shortly after. Thus, although the tangential collision curve itself does not signal instability, this boundary represents a useful proxy for dynamical lifetime, as objects that cross this curve are destabilized in tens of millions of years, a timescale much shorter than the lifetime of the solar system.

As a consequence of this evolution, the only objects which find long-term stability within the orbit-crossing region are ones trapped in mean-motion resonances with Planet Nine. Interestingly, all of the aligned objects (pink) whose longitude of perihelion librations protrude into the orbit-crossing region in our simulations are in fact in such resonances. The most visible ones are those objects at $a=700$ AU ($1$:$1$ resonance) and at $a=920$ AU ($2$:$3$ resonance). Such resonances are nevertheless rare, meaning that Planet Nine carves out a high-$q$ aligned population due to primarily secular effects.

\begin{figure}[t!]
\epsscale{1}
  \begin{center}
      \leavevmode
\includegraphics[width=85mm]{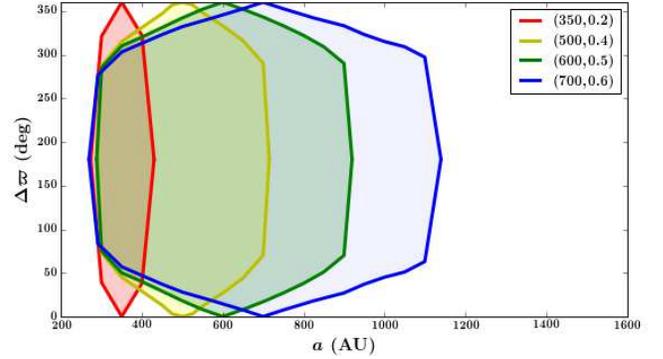}
\caption{Tangential collision curves enclosing the orbital crossing region for several different values of the Planet Nine orbital elements with a constant $q_9$. The shaded regions indicate the area of the $\Delta \varpi-a$ phase plane in which all orbital configurations of test particles found in the same plane as Planet Nine cross its orbit.} 
\label{fig:collision}
\end{center}
\end{figure}

We construct similar tangential collision curves for different Planet Nine configurations in Figure \ref{fig:collision}. Approximately fixing $q_9$, it is easy to see that lowering the semi-major axis and eccentricity of Planet Nine results in a shrinking orbital crossing region. Thus, closer in and more circular Planet Nine's would carve out the aligned and anti-aligned populations for a smaller range of test particle semi-major axes.

While our procedure yields well-defined vertical borders on the $a < a_9$ and  $a > a_9$ sides of Figure \ref{fig:collision}, it is important to note that the left and right boundaries of these orbit-crossing regions may not be physically meaningful for the aligned population. In the right panel of Figure \ref{fig:aligned}, we still find aligned objects at semi-major axis values of $a < 300$ AU and $a > 1100$ AU. Outside of this region, however, we are now in principle able to generate a population that circulates in $\Delta \varpi$ that does not require resonance for stability. In the $a < 300$ AU region, such objects have low eccentricities and aphelion distances less than the perihelion distance of Planet Nine. Similarly, in the $a > 1100$ AU region, this non-resonant circulating population is composed of objects with perihelion distances greater than Planet Nine's aphelion. In contrast, circulating objects in the orbit-crossing regime must reside in resonances to attain stability.

\subsection{The Inclined Model}

The results described thus far were found in our simplified, semi-averaged simulations. In order to verify that the computed dynamics is robust, we considered a more realistic numerical experiment. In this set of simulations, we ran test particles distributed in the same manner as above, but with a uniform inclination distribution $i \in [0^\circ, 40^\circ]$, and in the presence of an inclined Planet Nine, with $a_9 = 700$ AU, $e_9 = 0.6$, $i_9 = 20^\circ$, and $\omega_9 = 110^\circ$. In addition, we treated the four giant planets as active bodies, foregoing the $J_2$ approximation made in our planar simulations, and appropriately reducing the time-step to $1$ year. Both the narrow and broad Kuiper Belt simulations contained about $10,000$ particles, as before.

\begin{figure}[t!]
\epsscale{1}
  \begin{center}
      \leavevmode
\includegraphics[width=90mm]{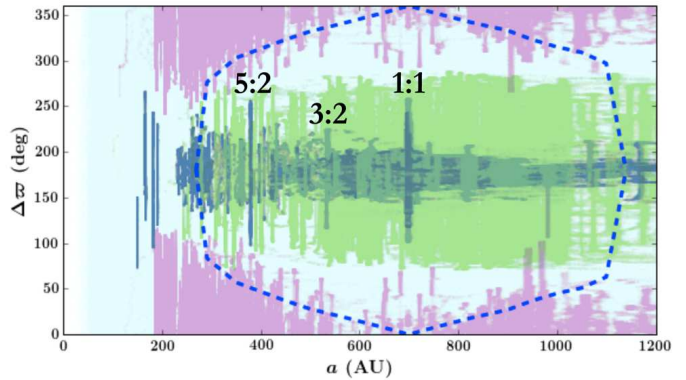}
\caption{The full trajectories of all surviving objects in the inclined simulations on the $\Delta \varpi-a$ phase plane. This surviving population includes objects that are aligned (pink), anti-aligned (dark blue), highly-inclined (green), and circulating in $\Delta \varpi$ (light blue). Although the inclined simulations are less ordered, the shape of the aligned population is similar to that shown in the right panel of Figure \ref{fig:aligned}. This behavior is created due to Planet Nine-induced secular dynamics.} 
\label{fig:inclined}
\end{center}
\end{figure}

Figure \ref{fig:inclined} shows the inclined simulation results on the $\Delta \varpi-a$ plane, and can be readily compared to the planar case shown in the right panel of Figure \ref{fig:aligned}. Clearly, the simulated particles exhibit a greater degree of stochasticity in their evolution; however, the general dynamical behavior is preserved. Interestingly, the libration width of the anti-aligned objects (blue) in Figure \ref{fig:inclined} appears to be smaller than that in the corresponding right panel of Figure \ref{fig:aligned}. A possible explanation for this is that due to modulation from Planet Nine's inclination, the edges of the anti-aligned orbital cluster breakup into a rapidly chaotic zone, reducing the width of the stable region \citep{BatMorb}. A more detailed analysis of this phenomenon would present an interesting avenue for future work.

Note that the dashed blue curve is the same planar collision curve as that shown in Figure \ref{fig:aligned}. The meaning of this boundary is now more complicated. Since Planet Nine and the test particles could have substantial mutual inclinations in these more realistic simulations, there are orbital configurations within the region bounded by this curve in which the orbits of the object and Planet Nine do not cross. Even so, due to the oscillations in inclination and longitude of ascending node, objects within the dashed curve predominantly reside in a domain in which close encounters are more likely. As a result, despite possessing prolonged dynamical lifetimes as compared to their counterparts in the strictly planar simulations, non-resonant objects found in this region will ultimately experience instability as well, and will be removed from the system.

\section{Discussion}
\label{sec:discussion}

In this work, we considered two synthetic Kuiper Belts with initially different perihelion distance distributions: a ``narrow" ($q_0 \in [30, 36]$ AU) and a ``broad" ($q_0 \in [30, 300]$ AU) population. Our numerical simulations show that after a $4$ Gyr integration, the final distribution of perihelion distance of these two populations remains distinct. Although both Kuiper Belt incarnations retain a low-$q$ anti-aligned population, only the broad distribution develops a high-$q$ aligned population. 

Qualitatively speaking, this discrepancy can be explained by Planet Nine-induced secular dynamics. The very high-$q$ anti-aligned objects do not survive the full integration time because they develop large librations and their trajectories are carried into the inner solar system. Analogously, the low-$q$ aligned objects are ejected once driven to $q < 30$ AU. Thus, we identify two \textit{permanently stable} populations in the outer Kuiper Belt: the low $q$ anti-aligned objects and the high-$q$ aligned objects.

While our initial simulations contained only a low inclination Kuiper Belt and a non-inclined Planet Nine, we ran a more realistic set of simulations, in which we considered an inclined Planet Nine and the effect of the giant planets as active bodies. Due to the good agreement between the results in the planar and inclined models, we are able to conclude that the planar approximation provides a reasonable representation of the dynamical behavior induced by Planet Nine. We can thus consider the implications of the results of our planar simulations as being representative.

Since different parameters of Planet Nine ``carve out" differing stable aligned populations, as shown in Figure \ref{fig:collision}, studying the distribution of $\Delta \varpi$ as a function of semi-major axis for distant Kuiper Belt objects as more are found presents an interesting constraint on the semi-major axis of Planet Nine. In the census of available observations, however, the sample of librating aligned objects is contaminated by objects that have a $\Delta \varpi \sim 0$, but are in fact circulating. The population of observable circulating objects dominates the observable aligned population, as the aligned objects are only stable at very high (and thus hard to observe) $q$ values. In light of this, a closer examination of the distribution of circulating objects may allow for stricter conclusions on the semi-major axis of Planet Nine, and presents an important avenue for future work.

More importantly, we find that the two incarnations of the distant Kuiper Belt are indeed distinct. Since all low-$q$ objects are depleted due to secular interactions with Planet Nine, the main difference between the narrow and broad Kuiper Belt distributions is in the existence of high-$q$ aligned objects. Thus, the presence of high-$q$ aligned objects in our solar system would not only lend support for the existence of Planet Nine but would indicate that our current Kuiper Belt stems from an initially wide $q$-distribution. This, in turn, would signal that Planet Nine's formation likely involved stellar encounters, as this is the most likely mechanism that would create such a spread out Kuiper Belt.

We note that the above analysis entails a caveat. In our model, we only consider the behavior of the outer Kuiper Belt; once we initialize the $\sim$10,000 particles, those are the only objects whose evolution we consider. That is, we do not take into account objects which may have originated in the inner Kuiper Belt or Scattered Disk, and later were scattered out to sufficiently high $a$, thus falling into our population of interest. These objects could replenish certain regions of parameter space, such as the low-$q$ aligned or the high-$q$ anti-aligned regions, adding an influx of objects to the existing populations. 

The objects that may be scattered into the outer solar system, however, will not become members of the two \textit{permanently stable} populations identified in this work (low-$q$ anti-aligned and high-$q$ aligned). Instead, these objects will replenish the orbital configurations that are eventually driven to centaur-like orbits and are then rapidly ejected from the solar system. As a result, these objects are only \textit{metastable}, as their lifetimes are generally shorter than one libration in $\Delta \varpi$, a process which lasts $ < 1$ Gyr. Even so, carefully modeling the flux of these scattered objects in the outer solar system under the influence of Planet Nine would be a possible direction of future research.

Under the assumption that our current and future observational census is dominated by the \textit{permanently stable} objects, we hope that upon discovery of additional distant KBOs, results identified in this study can be used to place meaningful constraints upon the initial distribution of the outer Kuiper Belt. In turn, the structure of this primordial population will allow us to draw conclusions regarding the formation pathway of Planet Nine, and by extension, the Sun's birth environment.

\textbf{Acknowledgements.} We thank Elizabeth Bailey, Christopher Spalding, and Juliette Becker for useful conversations, and Fred Adams for a careful review of the draft. We thank the anonymous referee for valuable comments that led to the improvement of the manuscript. T.K. is grateful to the Johnson and Johnson WAVE Fellowship for funding as well as to the Caltech Student-Faculty Program for their support of this work. K.B. acknowledges the generous support of the David and Lucile Packard Foundation.

\bibliographystyle{apj}
\bibliography{ref}

\end{document}